\documentclass[prd,onecolumn]{revtex4}
\usepackage{dcolumn}
\usepackage{multirow}
\usepackage{graphicx}
\usepackage{amssymb}
\usepackage{bm}
\usepackage{hyperref}
\usepackage{epstopdf}
\usepackage{color}
\usepackage{mathrsfs}
\usepackage{amsmath,amssymb,amsthm}
\begin{document}
\title{Wormholes from cosmological reconstruction based on Gaussian processes}
\author{Deng Wang}
\email{Cstar@mail.nankai.edu.cn}
\affiliation{Theoretical Physics Division, Chern Institute of Mathematics, Nankai University,
Tianjin 300071, China}
\author{Xin-He Meng}
\email{xhm@nankai.edu.cn}
\affiliation{Department of Physics, Nankai University, Tianjin 300071, China}
\begin{abstract}
We study the model-independent traversable wormholes from cosmological reconstruction based on Gaussian processes (GP). Using a combination of Union 2.1 SNe Ia data, the latest observational Hubble parameter data and recent Planck's shift parameter, we find that our GP method can give a tighter constraint on the normalized comoving distance, its derivatives and the dark energy equation of state than the previous work \cite{1}. Subsequently, two specific traversable wormhole solutions are obtained, i.e., the cases of a constant redshift function and a linear shape function. We find that, with decreasing cosmic acceleration, the traversal velocity $v$ of the former case increases and the amounts of exotic matter $I_V$ of the latter case decreases.
\end{abstract}
\maketitle
\section{Introduction}
Modern cosmology has already entered a precise, data-driven era. In 1998, the elegant discovery that the universe is undergoing the phase of accelerated expansion \cite{2,3}, has motivated a great deal of studies concentrated on how best to parameterize the dark energy and measure its properties. An accompanying task comes naturally into being, i.e., explore the corresponding behaviors at astrophysical scales when investigating the expansion history of the universe for different cosmological theories at cosmological scales. Recently, because of the discovery of the accelerating universe, theorists have gradually paid more and more attention to the exotic spacetime configurations, especially, the renewed field---wormholes. To be more concrete, in both cases (accelerating universe and wormholes) the null energy condition (NEC) is violated and consequently all of other energy conditions. Thus, an appealing overlap between two seemingly separated subjects occurs. So far, there is no doubt that together with black holes, white dwarfs, pulsars and quasars, etc., wormholes constituting the most intriguing celestial bodies may provide a new window for physical discovery.

In the literature, all the authors almost study the wormhole solutions only for a given dark energy model. However, with booming astronomical data, one can investigate better the wormhole spacetime configurations supported by cosmological observations through implementing highly precise constraints on a given dark energy model, in order to avoid choosing the values of the cosmological parameters arbitrarily. In our previous works \cite{4,5}, we have studied geometrical and holographical dark energy wormholes constrained by astrophysical observations for details, and verified that the exotic spacetime configurations wormholes can actually exist in our universe. It is worth noting that the obtained wormhole configurations are static and spherically symmetrical solutions existing at some certain low redshift. Nonetheless, the previous results depend apparently on given dark energy models. Hence, one shall ask whether the model-independent wormholes exist in the universe. In other words, whether can one investigate the existence of wormholes directly starting from cosmological observations ?

To answer this question, one should determine the dark energy equation of state (EoS) $\omega$ by using the model-independent methods, in order to satisfy the requirement $\omega<-1$ which ensures the wormhole spacetime structure open. During the past few years, there were many model-independent methods to study the evolutional behaviors of dark energy EoS $\omega$ in the literature, for instance, principal component analysis (PCA) \cite{6} and Gaussian processes (GP) \cite{1}, etc.

In this work, we would like to use the GP method to reconstruct the dark energy EoS $\omega$ and investigate the correspondingly model-independent wormhole solutions. The GP algorithm is a fully Bayesian approach for smoothing data, and can preform a reconstruction of a function directly from data without assuming a parameterization of the function. As a result, one can determine any cosmological quantity from cosmic data, and the key requirement of the GP algorithm is just the covariance function which entirely depends on the cosmic data. In light of the special advantage, the GP method has been widely applied in exploring the expansion dynamics of the universe \cite{1,7,8}, the test of the base cosmological model \cite{9}, the cosmography \cite{10}, the distance duality relation \cite{11}, the determination of the interaction between dark energy and dark matter and cosmic curvature \cite{12,13}, dodging the cosmic curvature to probe the constancy of the speed of light \cite{14}, dodging the matter degeneracy to determine the dynamics of dark energy \cite{15}, the slowing down of cosmic acceleration \cite{16}, etc. It is noteworthy that the analysis in paper \cite{17} has indicated that the Mat\'{e}rn ($\nu=9/2$) covariance function is a better choice to carry out the reconstruction and it has been used in papers \cite{9,12}.

The outline of the rest paper is as follows: In the next section, we make a brief review on GP method. In Section III, we reconstruct the dark energy EoS $\omega$ in order to study the corresponding wormhole spacetime configurations. In Section IV, we obtain two specific wormhole solutions including the cases of a constant redshift function and a linear shape function, and investigate the related physical properties. The discussions and conclusions are presented in the final section (we use the units $8\pi G=c=1$).

\section{GP method}
In a spatially flat Friedmann-Robertson-Walker (FRW) universe, the luminosity distance $d_L(z)$ can be expressed as
\begin{equation}
d_L(z)=\frac{1+z}{H_0}\int^{z}_{0}\frac{dz'}{E(z')}, \label{1}
\end{equation}
where $z$ denotes the redshift, $H_0$ the present-day value of the Hubble parameter and $E(z)$ the dimensionless Hubble parameter, respectively. Subsequently, using the normalized comoving distance $D(z)=H_0(1+z)^{-1}d_L(z)$, the dark energy EoS is written as
\begin{equation}
\omega(z)=\frac{2(1+z)D''-3D'}{3D'[(1+z)^3\Omega_{m0}D'^2-1]}, \label{2}
\end{equation}
where the prime denotes the derivative with respect to (w.r.t) the redshift $z$ and $\Omega_{m0}$ is present-day value of the matter density ratio parameter.

To study the model-independent wormhole configurations, firstly, one needs to reconstruct the dynamic dark energy EoS $\omega(z)$. Then, as described in our previous works \cite{4,5,18,19}, one shall concentrate on the parts in which the condition $\omega(z)<-1$ is satisfied and study the static and spherically symmetrical wormhole solutions at some certain redshift.

The GP method can reconstruct a function directly from observational data without assuming a concrete parameterization for the function. Here we utilize the package GaPP (Gaussian Processes in Python) \cite{1} to implement the reconstruction, which is firstly invented by Seikel et al. for a pedagogical introduction to GP.   Usually, the GP is a generalization of a Gaussian distribution which is the distribution of a random variable. In addition, the GP exhibits a distribution over functions. At each point $x$, the reconstructed function $f(x)$ is a Gaussian distribution with a mean value and Gaussian error. The key of the GP method is a covariance function $k(x,\tilde{x})$ which correlates the function $f(x)$ at different reconstruction points. The covariance function $k(x,\tilde{x})$ depends entirely on two hyperparameters $l$ and $\sigma_f$, which characterize the coherent scale of the correlation in $x$-direction and typical change in the $y$-direction, respectively. In general, the choice is the squared exponential covariance function $k(x,\tilde{x})=\sigma_f^2 \mathrm{exp}[-|x-\tilde{x}|^2/(2l^2)]$. However, the analysis in \cite{17} has verified that the Mat\'{e}rn ($\nu=9/2$) covariance function is a better choice to carry out the reconstruction. Hence, we will adopt the Mat\'{e}rn ($\nu=9/2$) covariance function in the following analysis:
\begin{equation}
k(x,\tilde{x})=\sigma_f^2 \mathrm{exp}(-\frac{3|x-\tilde{x}|}{l})\times[1+\frac{3|x-\tilde{x}|}{l}+\frac{27(x-\tilde{x})^2}{7l^2}+\frac{18|x-\tilde{x}|^3}{7l^3}+\frac{27(x-\tilde{x})^4}{35l^2}]. . \label{3}
\end{equation}

\begin{figure}
\centering
\includegraphics[scale=0.6]{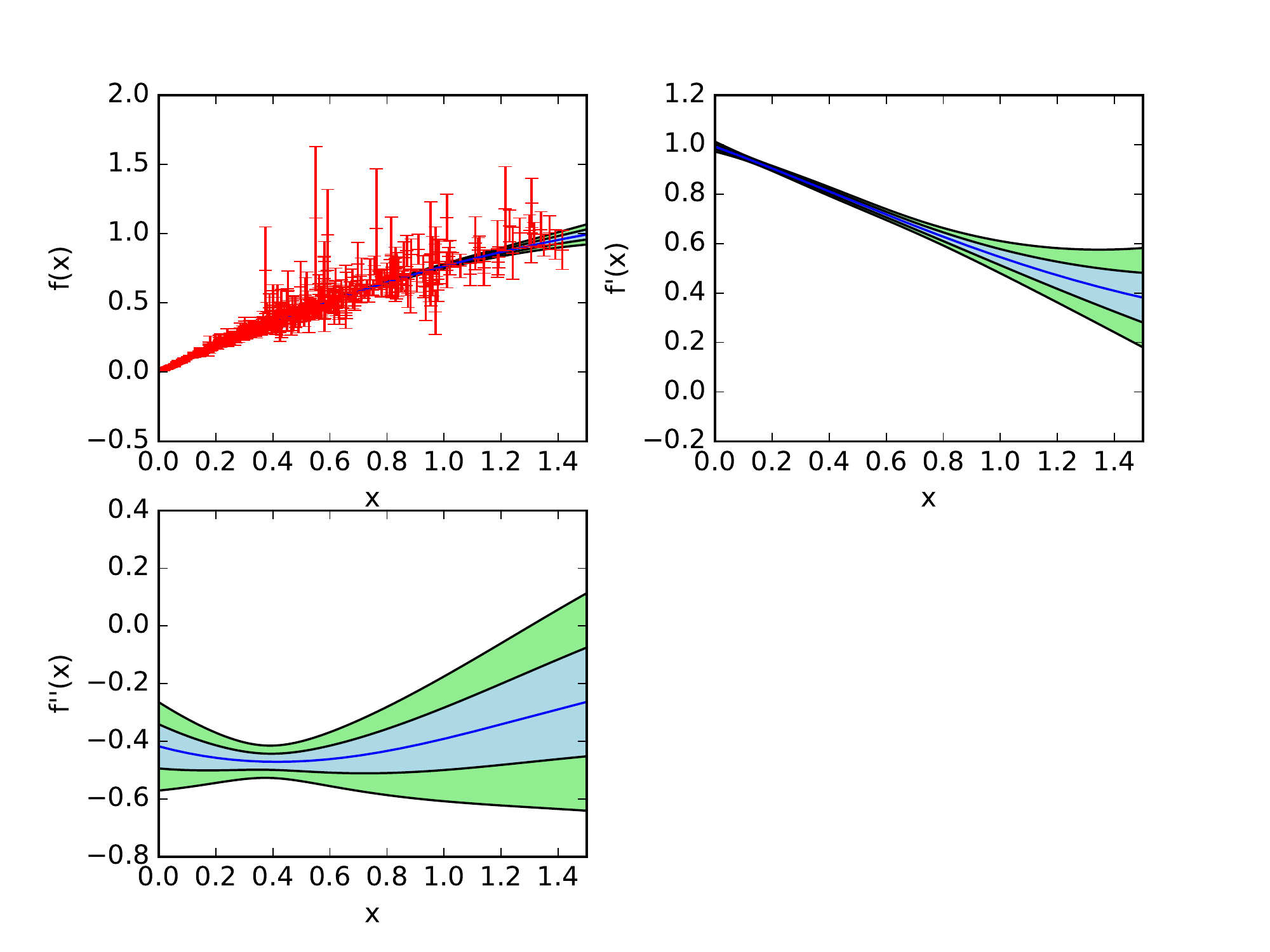}
\caption{The GP reconstruction of $f(x), f'(x)$ and $f''(x)$ using the Union 2.1 data sets. The shaded regions are reconstructions with $68\%$ and $95\%$ confidence level. The blue lines represents the underlying true model (the mean value of the reconstructions). Since this plot is aimed at verifying the correctness of our GP method, we take the labels $x, f(x), f'(x)$ and $f''(x)$ as distinctions with those $z, D(z), D'(z)$ and $D''(z)$ used in Fig. \ref{f2}.}\label{f1}
\end{figure}

In our reconstruction, we use the Union 2.1 data sets \cite{20} which consist of 580 SNe Ia data and cover the redshift range [0.015,1.4]. As noted in paper \cite{1}, we transform the distance modulus $m-M$ to $D$ in the following manner
\begin{equation}
m-M-25+5\lg(\frac{H_0}{c})=5\lg[(1+z)D] \label{4}
\end{equation}
with $H_0=70kms^{-1}Mpc^{-1}$. Furthermore, we set the initial conditions $D(z=0)$ and $D'(z=0)=1$ in the reconstruction process. Notice that the values of $D$ just depend on a combination of the absolute magnitude $M$ and $H_0$. Different from the previous literature \cite{1,7,8,9,10,11,12,13,14,15,16,17}, we adopt the latest 36 observational Hubble parameter data (OHD) from the paper \cite{21} and the shift parameter $\mathcal{R}=1.7488\pm0.0074$ from the recent Planck's result \cite{22} as important supplements of SNe Ia data, since the GP algorithm in GaPP does need the data of $D$ and $D'$ at the same time. To be more precise, we conclude the relations between the above-mentioned data sets and the normalized comoving distance $D$ as follows
\begin{equation}
m-M\Longrightarrow D,\label{5}
\end{equation}
\begin{equation}
\frac{H_0}{H(z)}\Longrightarrow D',\label{6}
\end{equation}
\begin{equation}
\mathcal{R}=\sqrt{\Omega_{m0}}\int^{z_c}_0\frac{dz'}{E(z')}\Longrightarrow D,\label{7}
\end{equation}
where $z_c=1089.0$ denotes the redshift of recombination and we use the values $H_0=70kms^{-1}Mpc^{-1}$ and $\Omega_{m0}=0.27\pm0.015$ \cite{20}. Subsequently, we verify the correctness of our method by using the Union 2.1 data sets to run the modified code. Obviously, the result in Fig. \ref{f1} is consistent with that in Fig. 6 of the paper \cite{16}.
\begin{figure}
\centering
\includegraphics[scale=0.6]{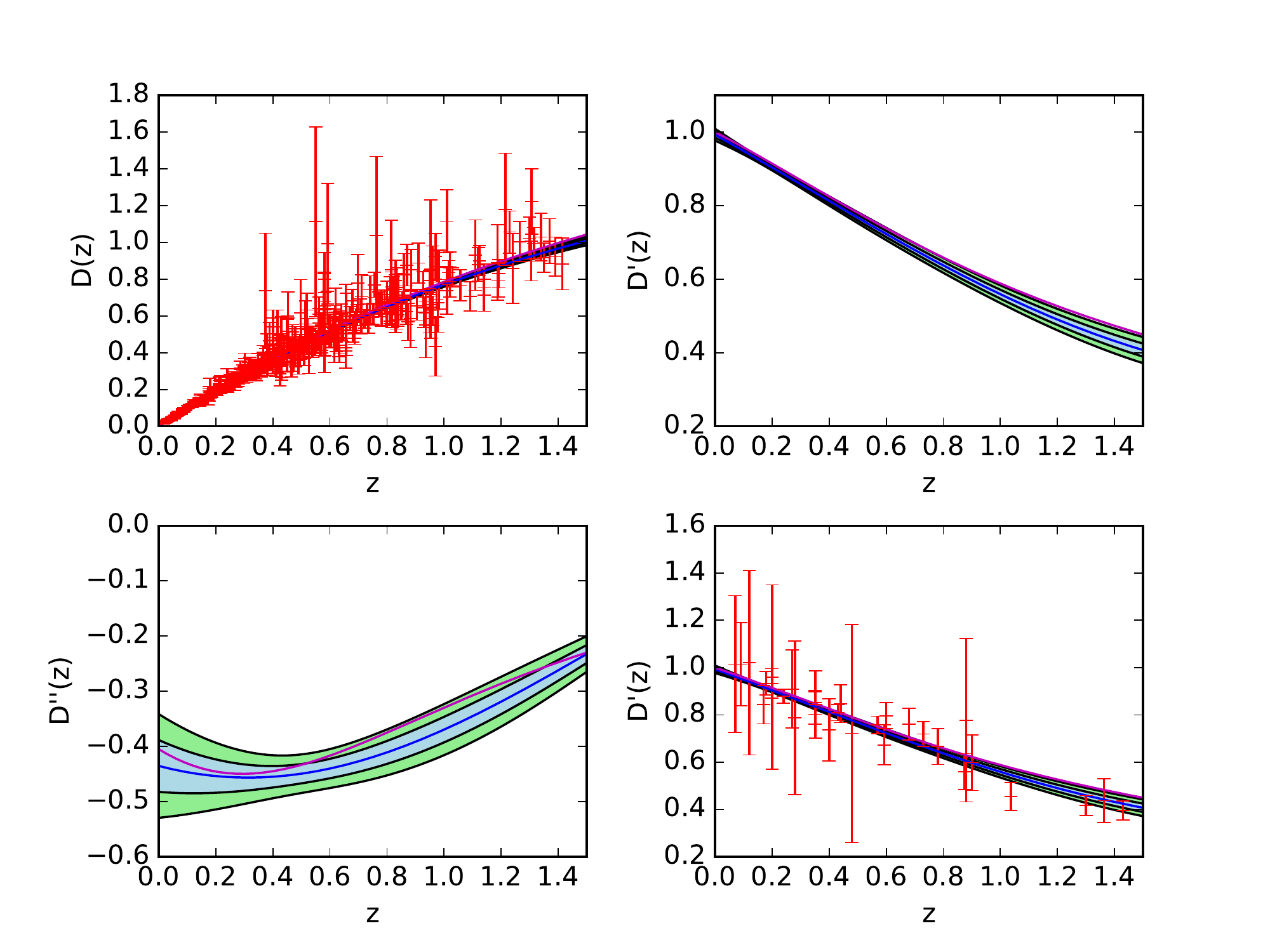}
\caption{The GP reconstruction of $D(z), D'(z)$ and $D''(z)$ using a combination of the Union 2.1 data sets, 36 OHD data points and Planck's shift parameter. The 36 OHD data points are shown in the lower right panel. The blue and magenta lines represents the underlying true model (the mean value of the reconstructions) and the base cosmological model, respectively.}\label{f2}
\end{figure}
\begin{figure}
\centering
\includegraphics[scale=0.3]{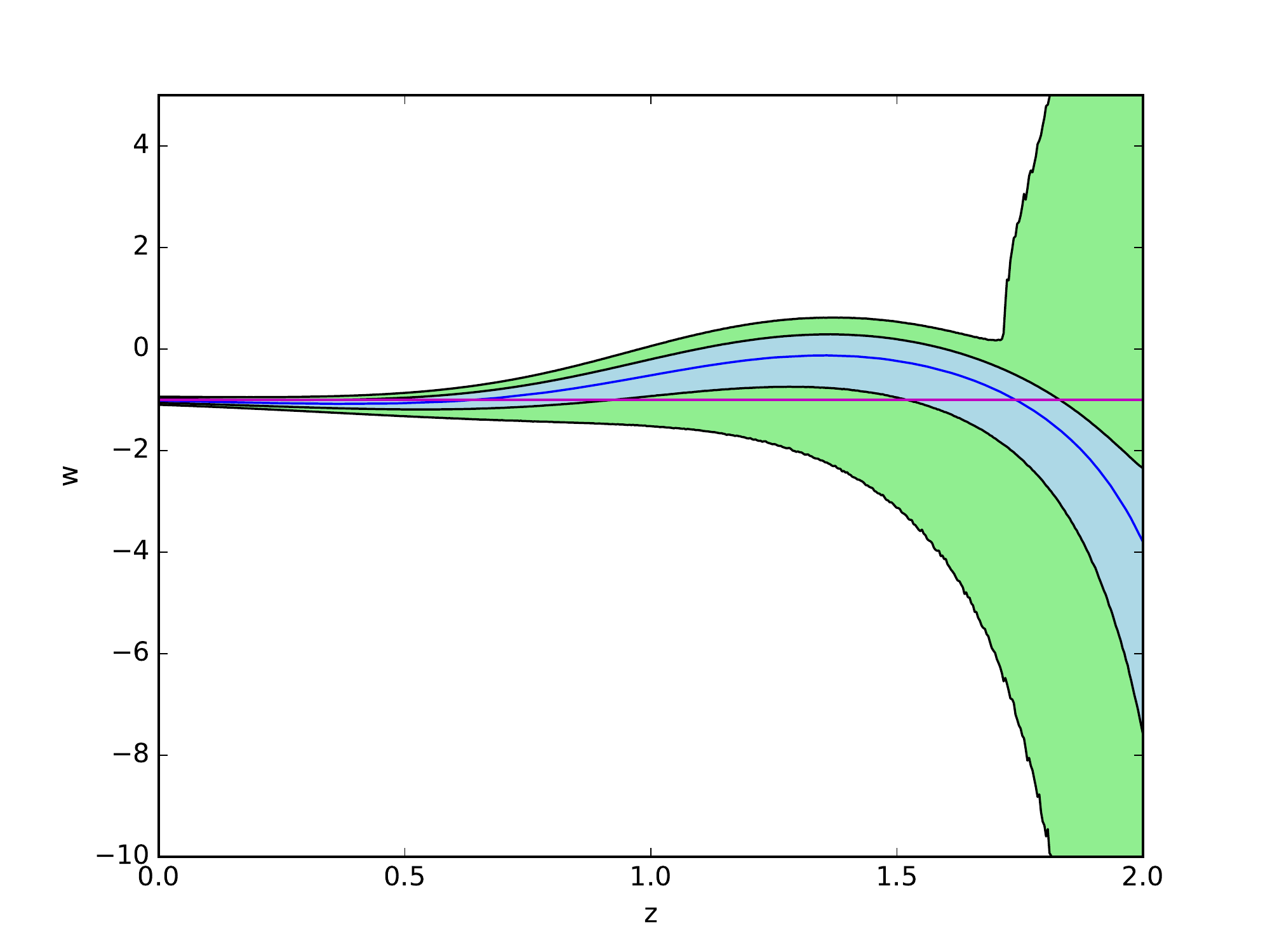}
\includegraphics[scale=0.3]{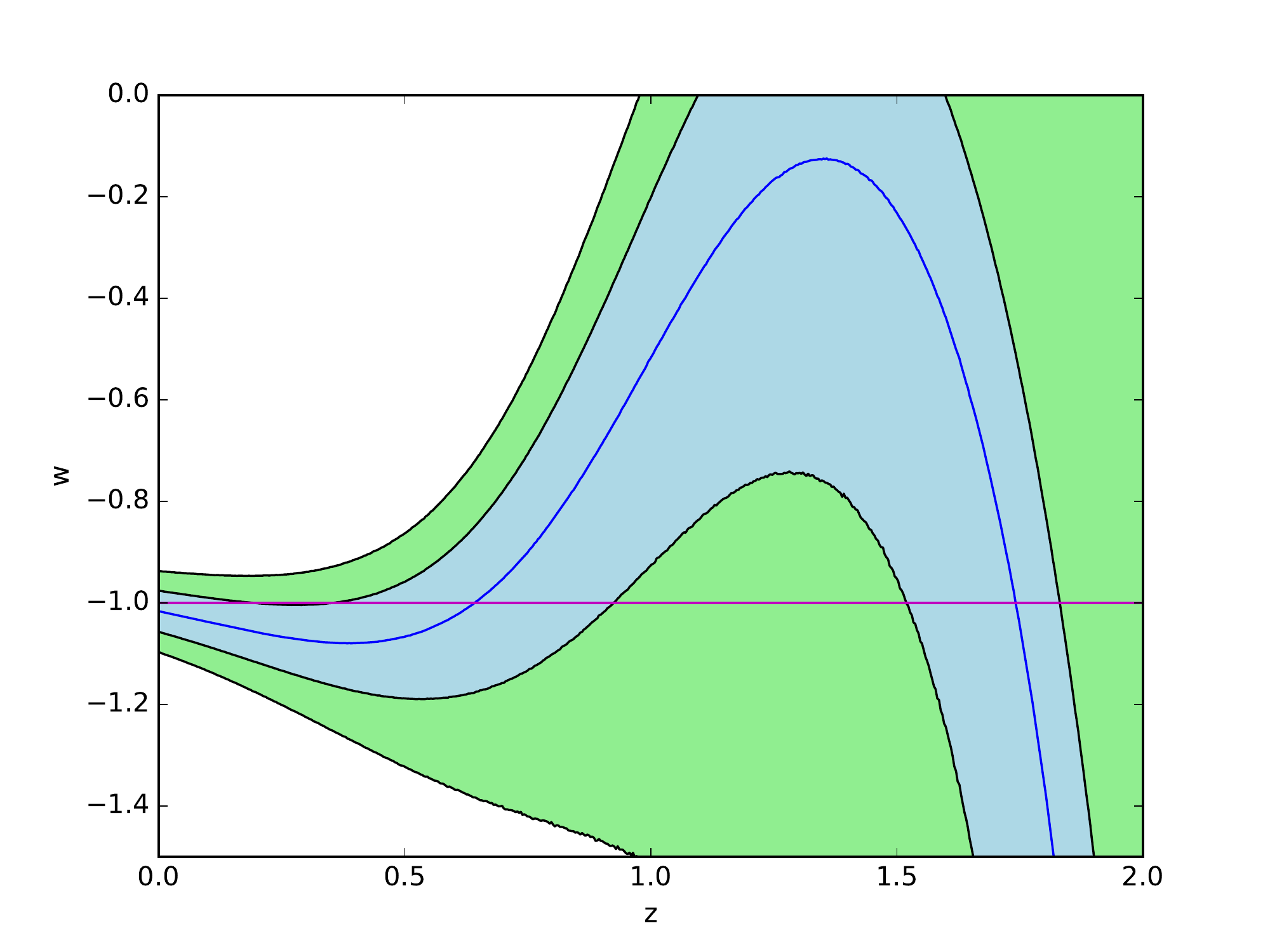}
\caption{The GP reconstruction of dark energy EoS $\omega(z)$ using a combination of the Union 2.1 data sets, 36 OHD data points and Planck's shift parameter. The blue and magenta lines represents the underlying true model (the mean value of the reconstructions) and the base cosmological model, respectively. The left and right panels represent the longitudinal coordinate ranges [-10,5] and [-1.5,0], respectively.}\label{f3}
\end{figure}
\section{Reconstruction results}
In this section, we implement the reconstruction by using a combination of Union 2.1 data sets, 36 OHD data points and Planck's shift parameter, and the result is shown in Fig. \ref{f2}. It is easy to find that the combined reconstructions $D(z), D'(z)$ and $D''(z)$ in Fig. \ref{f2} give apparently a tighter constraint  than those in Fig. \ref{f1}, and that the base cosmological model is consistent with the GP reconstructions at $2\sigma$ level. More precisely, the OHD data provide a tighter constraint in the low redshift range, and the Planck's shift parameter gives a substantially strict high-redshift constraint avoiding the divergence in the high-redshift range of Fig. \ref{f1}. Therefore, we can reconstruct the dark energy EoS $\omega(z)$ better in light of the stricter constraints on $D(z), D'(z)$ and $D''(z)$. Then, according to Eq. (\ref{2}), we carry out the reconstruction in Fig. \ref{f3}. To exhibit better the evolving behaviors of $\omega(z)$, we make two plots in the longitudinal coordinate ranges [-10,5] and [-1.5,0], respectively. We find that we give a tighter constraint on $\omega(z)$ by comparing Fig. \ref{f3} with Fig. 9 in paper \cite{1}. Furthermore, we find that the base cosmological model is well compatible with our reconstruction at $2\sigma$ level, and that the dark energy EoS $\omega(z)$ may exhibit an oscillating behavior at $1\sigma$ level in the low redshift range [0, 2]. Concentrating on the parts $\omega<-1$ in Fig. \ref{f3}, this result implies that the evolving wormhole spacetime configurations supported by cosmological observations may actually exist in the universe. However, in the present paper, we just investigate the static and spherically symmetrical wormhole configurations at some certain redshift.

\section{Wormhole models}
Consider a general metric for the wormhole models as follows
\begin{equation}
ds^2=-e^{2\Phi(r)}dt^2+\frac{dr^2}{1-\frac{b(r)}{r}}+r^2(d\theta^2+\sin^2\theta d\phi^2), \label{8}
\end{equation}
where $r$, $\theta$, $\phi$, $\phi(r)$ and $b(r)$ denote the radial coordinate, angular coordinates, the redshift function and shape function, respectively. Usually speaking, there are four necessary ingredients to form a wormhole configuration \cite{23}:

$\diamond$ The NEC is violated.

$\diamond$ Satisfy the so-called flaring-out conditions, i.e., $b(r_0)=r_0$, $b'(r_0)<1$ and $b(r)<r$ when $r>r_0$.

$\diamond$ $\Phi(r)$ shall be finite anywhere so that an horizon can be avoided.

$\diamond$ The asymptotically flatness condition shall be satisfied, i.e., $b/r\rightarrow0$ and $\Phi\rightarrow0$ when $r\rightarrow\infty$.

In an orthonormal reference frame, using the Einstein field equation, $G_{\mu\nu}=T_{\mu\nu}$, we express the stress-energy scenario as:

\begin{equation}
\rho=\frac{b'}{r^2}, \label{9}
\end{equation}
\begin{equation}
p_r=\frac{b}{r^3}-2\frac{\Phi'}{r}(1-\frac{b}{r}), \label{10}
\end{equation}
\begin{equation}
p_t=(1-\frac{b}{r})[\Phi''+(\Phi')^2-\Phi'\frac{b'r-b}{2r^2(1-b/r)}+\frac{\Phi'}{r}-\frac{b'r-b}{2r^3(1-b/r)}], \label{11}
\end{equation}
where $\rho(r)$, $p_r(r)$, $p_t(r)$ and the prime denote the matter energy density, the radial pressure, the tangential pressure and the derivative w.r.t. $r$, respectively. Using the stress-energy conservation equation, $T^{\mu\nu}_{\hspace{3mm};\nu} = 0$, one can have
\begin{equation}
p'_r=\frac{2}{r}(p_t-p_r)-\Phi'(\rho+p_r), \label{12}
\end{equation}
which can be regarded as the relativistic Euler equation or the hydrostatic equation of equilibrium for the matter threading a wormhole spacetime structure. Subsequently, we would like to calculate the wormhole solutions by adopting the mean value $\omega_{m}=-1.075$, which lies in the $1\sigma$ range $\omega(0.5)\in[-1.188, -0.982]$ at redshift $z=0.5$ (see Fig. \ref{f3}).

\subsection{A Constant Redshift Function}
Take into account a constant redshift function $\Phi=C$, substituting the dark energy EoS $p=\omega_{m}\rho$ and Eq. (\ref{9}) into Eq. (\ref{10}), we obtain
\begin{equation}
b(r)=r_0(\frac{r_0}{r})^{\frac{1}{\omega_{m}}}. \label{13}
\end{equation}
One can easily check that this shape function $b(r)<r$ when $r>r_0$, which satisfies the flaring-out conditions. Subsequently, evaluating at the wormhole throat $r_0$, we get $b'(r_0)=-\frac{1}{\omega_{m}}$. Furthermore, taking the mean value $\omega_{m}=-1.075$ from GP reconstruction, one can also get $b'(r_0)=-0.93<1$. Because the constant redshift function $\Phi$ is finite everywhere and $b/r\rightarrow0$ when $r\rightarrow\infty$, this wormhole spacetime structure is both asymptotically flat and traversable. Thus, the dimension of this wormhole configuration can be substantially large in principle.

In what follows, we would like to analyze the the traversability of the wormhole configuration as our previous works \cite{4,5,18,19}. Generally speaking, for a traveler in the spaceship who plans to journey successfully through the wormhole throat, there are three necessary ingredients [23]:

$\diamond$ The tidal acceleration shall not exceed 1 Earth's gravitational acceleration $g_\star$.

$\diamond$ The acceleration felt by the travelers shall not exceed 1 Earth's gravitational acceleration $g_\star$.

$\diamond$ The traverse time measured by the travelers and the observers who keep static at the space station shall satisfy the quantitative relations.

More detailed descriptions can be found in paper \cite{23}. For simplicity, we only exhibit the key formula through some derivations in the following manner
\begin{equation}
v\leqslant r_0\sqrt{\frac{\omega_{m} g_\star}{\omega_{m}+1}}, \label{14}
\end{equation}
\begin{figure}
\centering
\includegraphics[scale=0.4]{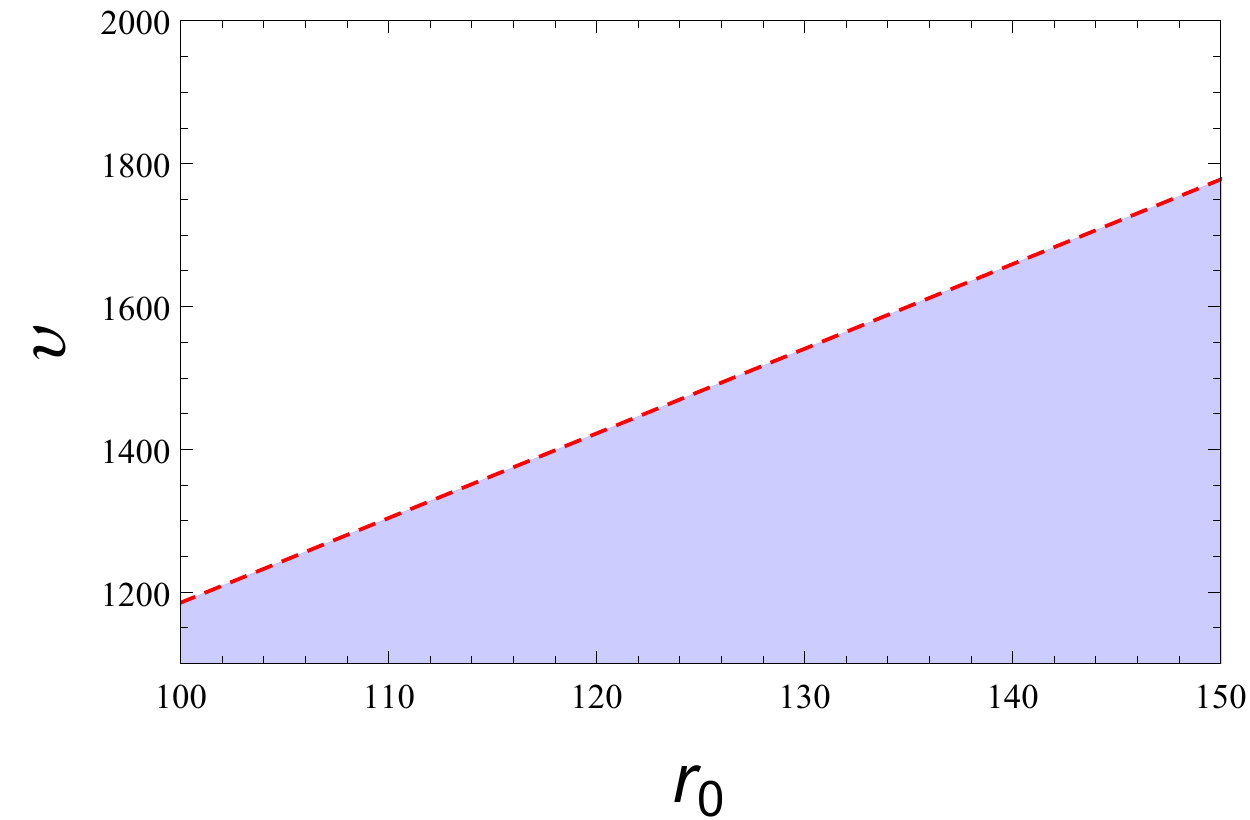}
\includegraphics[scale=0.4]{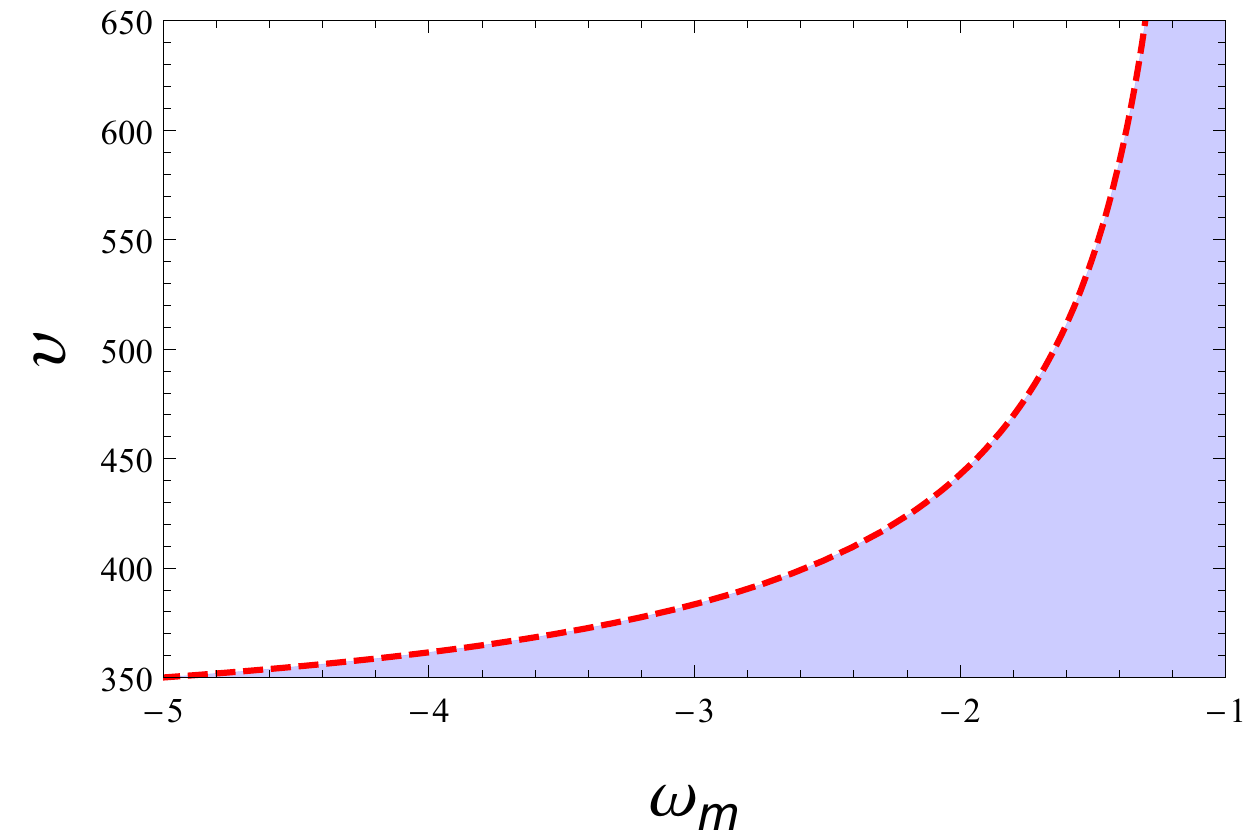}
\includegraphics[scale=0.4]{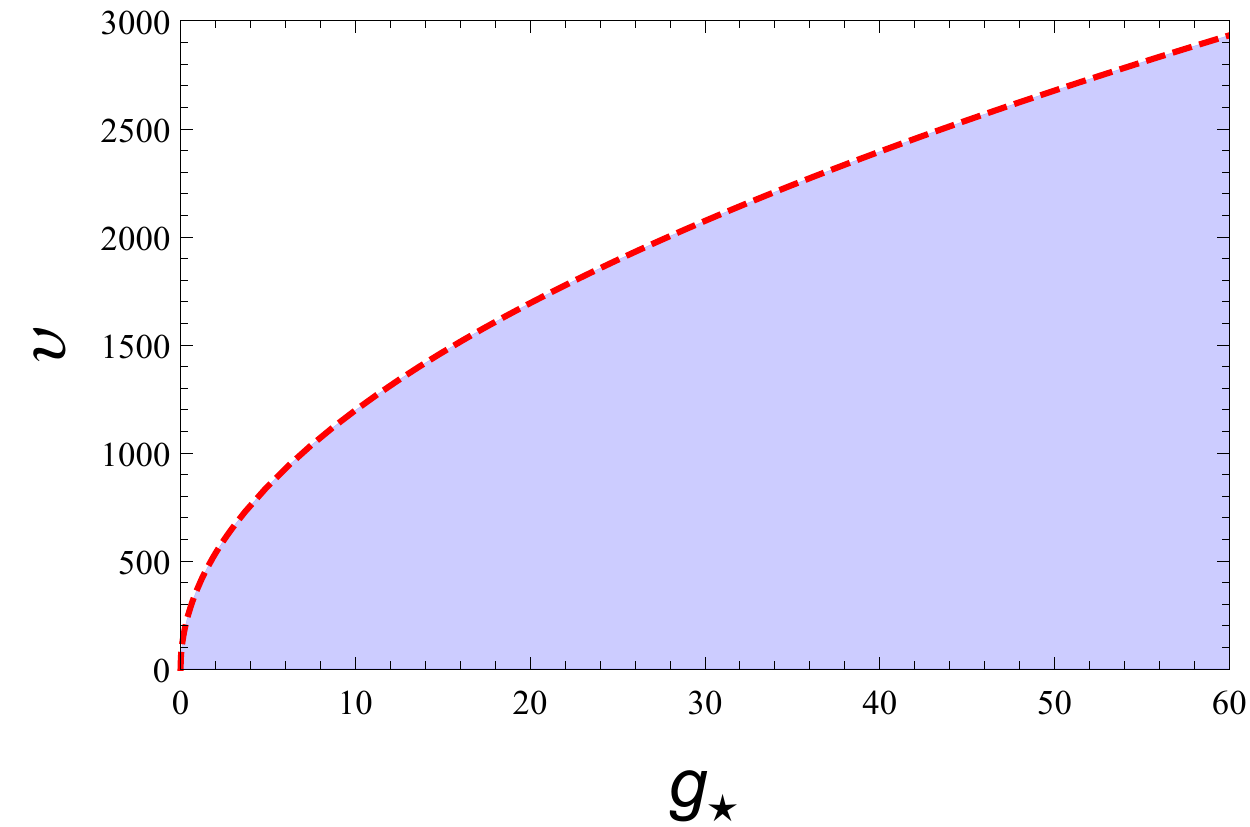}
\caption{In the left panel, we exhibit the relation between the traversal velocity $v$ and the throat radius of the wormhole $r_0$ by assuming $\omega_{m}=-1.075$ and $g_\star=9.8$ m$/$s$^{2}$. In the medium panel, we exhibit the relation between the traversal velocity $v$ and the dark energy EoS $\omega_{m}$ by assuming $r_0=100$ m and $g_\star=9.8$ m$/$s$^{2}$. In the right panel, we exhibit the relation between the traversal velocity $v$ and the Earth's gravitational acceleration $g_\star$ by assuming $r_0=100$ m and $\omega_{m}=-1.075$. The shaded regions and the red (dashed) lines correspond to the allowable regions for the traversal velocity $v$ and the extreme cases the equality signs in inequality (\ref{14}). }\label{f4}
\end{figure}
where $v$ is the traversal velocity. Subsequently, assuming the height of the traverser $h=2$ m, $r_0=100$ m, $\omega_{m}=-1.075$ and $g_\star=9.8$ m$/$s$^{2}$, we can have the traversal velocity $v\approx1185.19$ m/s, which is approximately equivalent to 3 times the speed of sound. Furthermore, we can also get the traverse time $\Delta\tau\approx\Delta t\approx2r_\star/v\thickapprox16.875s$ by setting the junction radius $r_\star=10000$ m. Moreover, one can find that the value of the traversal velocity $v$ in inequality (\ref{14}) depends on the throat radius of the wormhole $r_0$, the dark energy EoS $\omega_{m}$ from the GP reconstruction and the Earth's gravitational acceleration $g_\star$. In addition, we are also of interest to investigate the dependence of the traversal velocity $v$ on the above-mentioned physical quantities.

In the left panel of Fig. \ref{f4}, it is easy to be seen that the traversal velocity $v$ is a linear function of the throat radius of the wormhole $r_0$, meaning that with increasing $r_0$, $v$ will also gradually increase. In the medium panel of Fig. \ref{f4}, we find that the traversal velocity $v$ increase monotonically with gradually dark energy EoS $\omega_{m}$, which means that $v$ will increase with decreasing cosmic acceleration. Notice that the dark energy EoS $\omega_{m}$ will not approach $-1$ since the right hand side of inequality (\ref{14}) diverges. In the right panel of Fig. \ref{f4}, assuming that the largest gravitational acceleration a human being can bear is 6.5 times Earth's gravitational acceleration, we find that the traversal velocity $v$ still increase monotonically with increasing gravitational acceleration $g_\star$.

\subsection{A Linear Shape Function}
Take into account a specific shape function $b(r)=r_0-\frac{1}{\omega_{m}}(r-r_0)$ and utilize the dark energy EoS $p=\omega_{m}\rho$, we have
\begin{equation}
\Phi'(r)=-\frac{1}{2r} \qquad and \qquad  \Phi(r)=-\frac{1}{2}\ln r+C, \label{15}
\end{equation}
where $C$ denotes an arbitrary integration constant. We find that this solution is non-asymptotically flat since it diverges directly when $r\rightarrow\infty$.
Thus, this solution is non-traversable for an interstellar traveler. However, in theory, one can construct a traversable wormhole geometry through gluing an exterior flat geometry into the interior geometry at a junction radius $r_\diamond$. So the constant $C$ can be written as $C=\Phi(r_\diamond)+\frac{1}{2}\ln(\frac{r_\diamond}{r})$.

\begin{figure}
\centering
\includegraphics[scale=0.4]{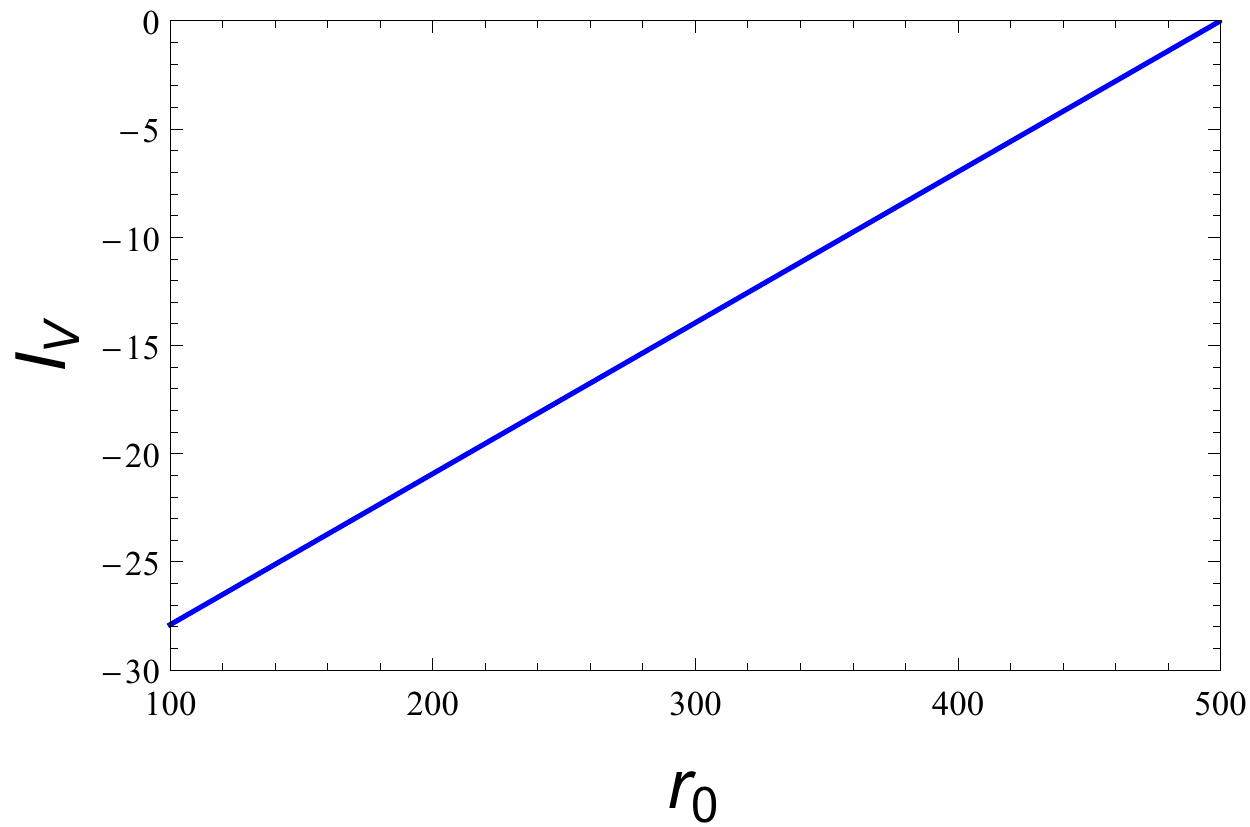}
\includegraphics[scale=0.4]{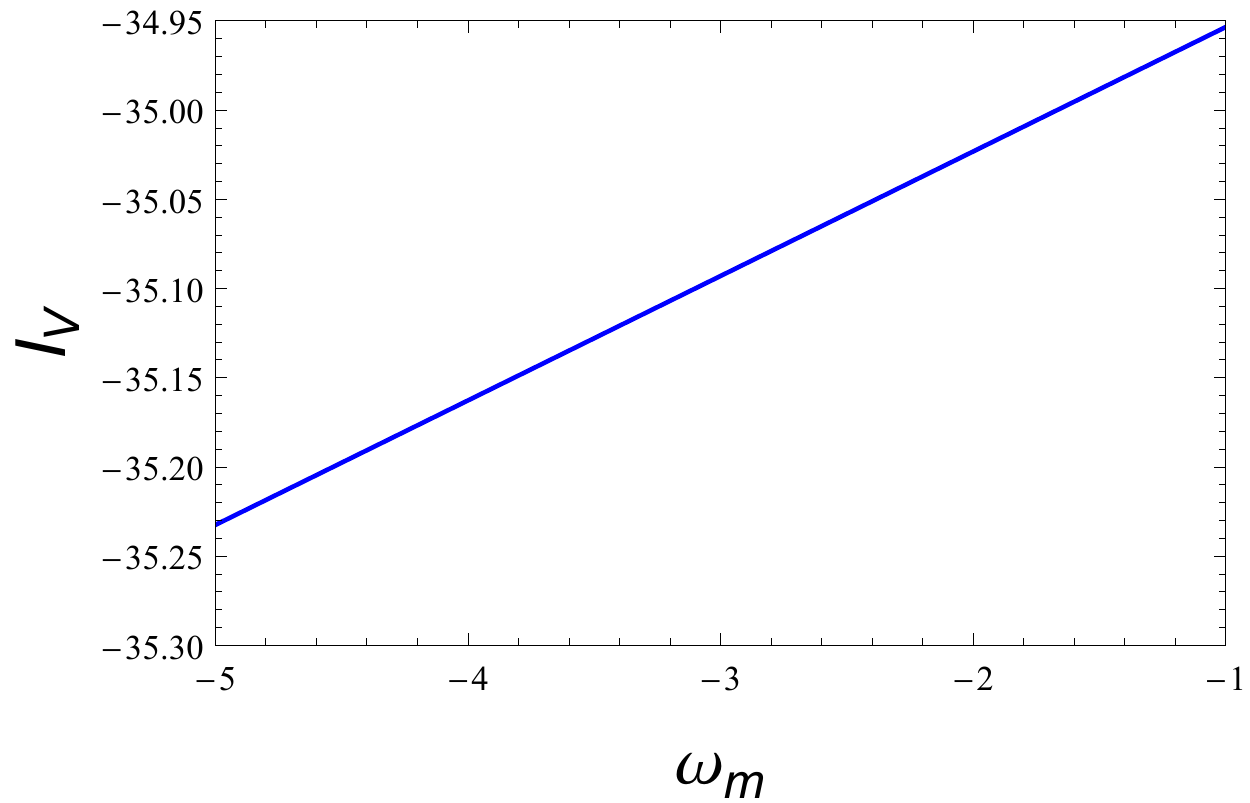}
\includegraphics[scale=0.4]{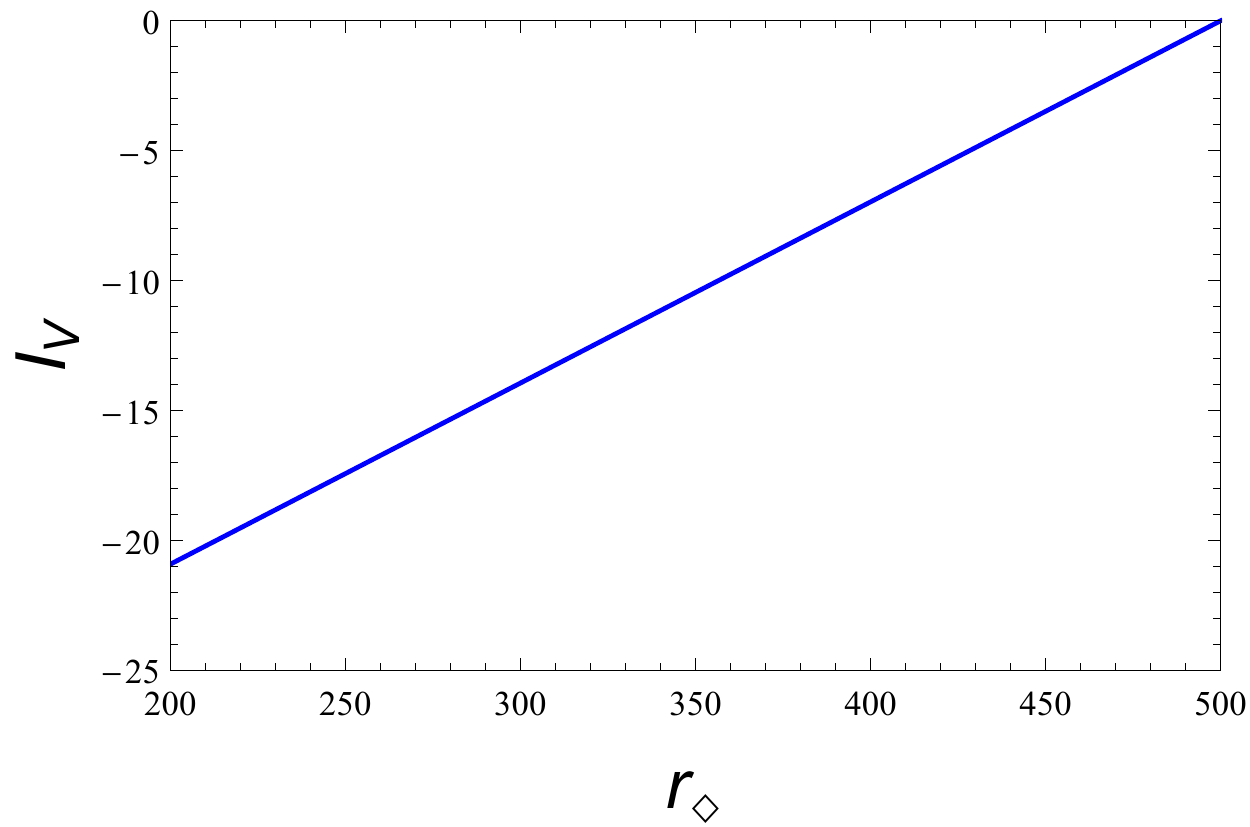}
\caption{In the left panel, we exhibit the relation between the amounts of exotic matter $I_V$ and the throat radius of the wormhole $r_0$ by assuming $\omega_{m}=-1.075$ and $r_\diamond=500$ m. In the medium panel, we exhibit the relation between the amounts of exotic matter $I_V$ and the dark energy EoS $\omega_{m}$ by assuming $r_0=100$ m and $r_\diamond=500$ m. In the right panel, we exhibit the relation between the amounts of exotic matter $I_V$ and the junction radius $r_\diamond$ by assuming $r_0=100$ m and $\omega_{m}=-1.075$. }\label{f5}
\end{figure}

Subsequently, we will utilize the the method of `` volume integral quantifier '' (VIQ) \cite{24} to quantify the exotic matter constructing the traversable wormhole in the finite range $r_0\leqslant r\leqslant r_\diamond$. For simplicity, one can obtain the amounts of the exotic matter by calculating the definite integral $\int T_{\mu\nu}k^\mu k^\nu dV$, where $T_{\mu\nu}$ is still the stress-energy tensor and $k^{\mu}$ any future directed null vector. In what follows, computing the quantity $I_V=\int[p_r(r)+\rho]dV$, we obtain
\begin{equation}
I_V=\int^{r_\diamond}_{r_0}(r-b)[\ln(\frac{e^{2\Phi}}{1-\frac{b}{r}})]'dr. \label{16}
\end{equation}
Through some easy derivations, we get
\begin{equation}
I_V=(1+\frac{1}{\omega_{m}})(r_0-r_\diamond). \label{17}
\end{equation}
Since the amounts of exotic matter $I_V$ is associated with the throat radius of the wormhole $r_0$, the dark energy EoS $\omega_{m}$ from the GP reconstruction and the junction radius $r_\diamond$, it is necessary to study the the dependence of the amounts of exotic matter $I_V$ on the above-mentioned physical quantities. In the left and right panels of Fig. \ref{f5}, we find that $I_V$ increases with increasing $r_0$ and $r_\diamond$ and approaches zero finally, since it is a linear function of $r_0$ and $r_\diamond$. In the medium panel of Fig. \ref{f5}, however, we find that $I_V$ increases very slowly with increasing $\omega_{m}$. This can be ascribed to the value of $r_0-r_\diamond$ is very large. It is not difficult to verify that the amounts of exotic matter $I_V\longrightarrow0$ when $r_j\rightarrow r_0$ and $\omega_{m}$ is fixed. For instance, if we still use the dark energy EoS $\omega_{m}=-1.075$ from GP reconstruction, Eq. (\ref{17}) can be rewritten as
\begin{equation}
I_V=0.0698(r_0-r_\diamond). \label{18}
\end{equation}
This result indicates that, in theory, one can construct a traversable wormhole with infinitesimal amounts of averaged-NEC (ANEC) violating dark energy fluid in the present situation. It is worth noting that we have obtained the same result by using the model-independent GP method, which does not depend on any concretely cosmological model.

\section{Discussions and conclusions}
The elegant discovery that the universe is in accelerating expansion has motivated theorists to pay afresh attention to the old field---wormholes, since in both fields an surprising overlap occurs, i.e., the violation of the NEC. However, in the literature, all the authors almost study the wormhole spacetime configurations for a given cosmological model. Therefore, in light of rapidly increasing cosmic data, we are aiming at exploring the model-independent dark energy wormholes by using the GP method.

In this work, first of all, we make a brief review on the GP method and demonstrate the correctness of our reconstruction method. Subsequently, using a combination of Union 2.1 580 SNe Ia data, the latest 36 OHD data points and recent Planck's shift parameter, we find that our GP method can give a tighter constraint on the normalized comoving distance $D(z)$, its derivatives $D'(z), D''(z)$ and the dark energy EoS $\omega(z)$ than the previous literature \cite{1}. In the mean while, we find that the base cosmological model is well compatible with our reconstruction at $2\sigma$ level. In what follows, two specific traversable wormhole solutions are obtained in the framework of a perfect fluid, i.e., the cases of a constant redshift function and a linear shape function. By choosing the mean value $\omega_{m}=-1.075$, which lies in the $1\sigma$ range $\omega(0.5)\in[-1.188, -0.982]$ at redshift $z=0.5$ (see Fig. \ref{f3}), we analyze the traversabilities of the former case and investigate the dependence of the traversal velocity $v$ on the throat radius of the wormhole $r_0$, the dark energy EoS $\omega_{m}$ from the GP reconstruction and the gravitational acceleration $g_\star$. Furthermore, we also calculate the amounts of exotic matter of the latter case and study the the dependence of the amounts of exotic matter $I_V$ on the throat radius of the wormhole $r_0$, the dark energy EoS $\omega_{m}$ from the GP reconstruction and the junction radius $r_\diamond$. We find that, with decreasing cosmic acceleration, the traversal velocity $v$ of the former case increases and the amounts of exotic matter $I_V$ of the latter case decreases (see Figs. \ref{f4}-\ref{f5}).

The dark energy EoS $\omega(z)$ from our GP reconstruction may exhibit an oscillating behavior at $1\sigma$ level in the low redshift range [0,2], which indicates that the evolving wormhole spacetime configurations supported by cosmological observations may actually exist in the universe. Furthermore, the evolving wormhole spacetime structure originated from cosmic acceleration may open at a relatively high redshift and close at a low redshift based on some unknown mechanism (e.g., quantum fluctuations).

According to the papers \cite{25,26}, the obtained wormhole configurations from our GP construction can be stable by choosing appropriately the values of the throat radius and junction radius. However, the interesting stability analysis is beyond the scopic of the present work.

We expect more and more high quality cosmic data can help the human beings to explore the nature of dark energy and dark matter.

\section{acknowledgements}
The author Xin-He Meng warmly thanks Professors Bharat Ratra and S. D. Odintsov for helpful feedbacks on astrophysics and cosmology. The author Deng Wang thanks Qi-Xiang Zou for beneficial discussions and programming. The authors acknowledge partial support from the National Science Foundation of China.

\end{document}